\documentclass[english,10pt,pra,aps,showpacs,twocolumn,amsmath,amssymb,superscriptaddress,
longbibliography,
floats]{revtex4-1}
\usepackage[T1]{fontenc}
\usepackage[latin9]{inputenc}
\usepackage{verbatim}
\usepackage{refstyle}
\usepackage{amsmath}
\usepackage{amssymb}
\usepackage{graphicx}
\usepackage{color}
\usepackage{bm}
\usepackage{hyperref}
\usepackage{times}
\makeatletter

\usepackage{braket}

\usepackage{babel}
\begin{document}

\title{Low-density expansions for the homogeneous dipolar Bose gas at zero temperature}

\author{Alexander Yu.~Cherny}
\affiliation{Bogoliubov Laboratory of Theoretical Physics, Joint Institute for Nuclear
Research, 141980, Dubna, Moscow region, Russia}
\affiliation{Center for Theoretical Physics of Complex Systems, Institute for Basic Science (IBS), Daejeon 34051, Republic of Korea}

\date{\today}

\begin{abstract}
The low-density expansions for the energy, chemical potential, and condensate depletion of the homogeneous dilute dipolar Bose gas are obtained by regularizing the dipole-dipole interaction at long distances. It is shown that the leading term, proportional to the density, allows a simple physical interpretation and consistently describes the thermodynamic stability of the system. The long-range asymptotics are obtained analytically for the normal and anomalous one-particle correlation functions and the pair distribution function. We discuss the properties of the two-body scattering with zero relative momentum for the dipole-dipole interaction, in particular, we derive the asymptotics of the wave function and a correction to the scattering length for small values of the dipolar range. We show how the density expansions can be derived within the Bogoliubov model of weakly interacting particles without any divergence from the assumption of universality of the expansions at low densities.
\end{abstract}
\maketitle

\section{Introduction}
\label{sec:intro}

Currently, the  Bose-Einstein condensates in dipolar Bose gases are quite intensively studied both experimentally and theoretically (see, e.g., the reviews \cite{Baranov08,Lahaye09,Baranov12,
Yukalov18} and the book \cite{Pitaevskii16:book}). The first realization of Bose-Einstein condensate in the magnetic dipolar gas of $^{52}$Cr was reported in Ref.~\cite{Griesmaier05}. The dipolar forces can be attractive or repulsive along the different directions of the relative coordinate, and such anisotropy can lead to new phenomena unrealisable for usual one-component Bose-Einstein condensates (BECs). For instance, quantum droplets, that is, atomic or molecule clusters stabilized by the quantum fluctuations, were observed recently for the dipole gases of Dy \cite{pfau16,pfau16n} and Er \cite{Chomaz16}.

Theoretically, the ground-state properties of homogeneous dilute dipolar gases seem to be the simplest to obtain, however, this is actually not the case. The problem is the long-range nature of the dipolar potential, which falls off as $1/r^3$ at large distances. The power exponent, equal to three, lies exactly at the threshold separating short- and long-range interactions in three dimensions. For the inhomogeneous gases, the problem is amenable to solution with the Gross-Pitaevskii equation, because the interaction energy is finite due to a rapid decrease of the condensate wave function in an external trap.

The presence of the anisotropy makes the problem rather complicated. There is a competition between the attractive and repulsive parts of the dipole-dipole interaction: the repulsion in real space is favourable to the Bose-Einstein condensation in momentum space, while the attraction makes the system prone to condensation in real space, which might cause a collapse.

Strictly speaking, the Bogoliubov theory \cite{Bogoliubov47} is applicable only for short-range interparticle interactions. The ground-state energy depends on the Fourier transform of the pairwise interaction potential at zero momentum. However, the Fourier transform of the dipole-dipole potential is not defined at $q=0$, since it depends on the momentum direction. Nevertheless, the Bogoliubov formalism was formally applied to dipolar gases in Ref.~\cite{Lima12}. The energy per particle contains two terms: the Hartree mean-field term, proportional to the density of particles $n$, and the Lee-Huang-Yang correction, proportional to $n^{3/2}$. The Lee-Huang-Yang correction was correctly calculated by the authors of Ref.~\cite{Lima12}. It is important for explaining the stability of droplets, mentioned above (see, e.g., Ref.~\cite{Wachtler16}). At the same time, the Hartree term depends on the angle between the dipoles and momentum, which is an unphysical result.

The aim of this paper is to fill the gap in the literature and find the correct density expansions for the ground-state energy and chemical potential. To this end, the dipole-dipole interaction potential should be appropriately presented as a limiting case of some short-range potential, for which the Bogoliubov model can be applied. This procedure we call the regularization. The main result of the paper is given by Eqs.~(\ref{expen}) and (\ref{mudipolar}) below. The leading term, proportional to the density, allows a simple physical interpretation and consistently describes the thermodynamic stability of the system. Besides, the analytical expressions are obtained for the normal $\langle\hat{\psi}^{\dag}(\bm{r}_1)\hat{\psi}(\bm{r}_2)\rangle$ and anomalous $\langle\hat{\psi}(\bm{r}_1)\hat{\psi}(\bm{r}_2)\rangle$ one-particle correlators and for the density-density correlator at large distances. The decaying parts of the correlators become strongly anisotropic and can change  sign when the dipolar part of the interaction is sufficiently large.

The Bogoliubov expression for the ground-state energy contains an ultraviolet divergence after substituting the effective pseudopotential. The nature of the divergence and how to get rid of it are rarely discussed in the literature. We consider this issue in more detail for instructive purposes. As a byproduct, we obtain the contribution of the dipole-dipole interaction into the scattering length for a weak dipolar potential with a cutoff at small distances.

The paper is organized in the following way. In the section \ref{sec:naive}, we recall the basic relations of the Bogoliubov model needed for obtaining the density expansions and construct explicitly the regularized dipole-dipole interaction with the screened Coulomb potential when the range of the screening infinitely grows. This allows us to obtain the expansions we are looking for and the interaction energy in the Gross-Pitaevskii functional. Further, the thermodynamic stability of the system is examined, a physical interpretation of the leading term in the expansion of the round-state energy is given, and the normal and anomalous correlators and the pair distribution function are calculated. In the next section, we consider the two-body scattering amplitude at zero relative momentum for the dipolar potential with the cutoff. The properties of the two-body scattering amplitude are discussed;  in particular, we derive the asymptotics of the wave function and the first two terms in the Born series. In Sec.~\ref{sec:universality}, we consider the nature of the divergence arising in the Bogoliubov model with the pseudopotential and suggest a recipe how to avoid this problem by using the universality of the energy expansion at low densities. In the Conclusion, we discuss the results obtained and further prospects.

\section{The low-density expansions for the dipolar gas}
\label{sec:naive}

\subsection{The ground-state energy and condensate depletion in the Bogoliubov model}
\label{sec:bog}

The Bogoliubov model \cite{Bogoliubov47} considers a homogeneous system of $N$ spinless bosons, occupying the volume $V$ and interacting with a weak potential $V(\bm{r})$, whose Fourier transform is equal to $V(\bm{q})$. The ground-state energy per particle $\varepsilon=E/N$ is given by
\begin{align}
\varepsilon =& \frac{n}{2}\left[V(0) - \int\frac{\mathrm{d}^3q}{(2\pi)^3}\frac{V^2(\bm{q})}{2T_{q}}\right]\nonumber\\ &+\frac{1}{2n}\int\frac{\mathrm{d}^3q}{(2\pi)^3} \left[\omega_{\bm{q}}-T_{q}-n V(\bm{q})+\frac{n^2V^2(\bm{q})}{2T_{q}}\right]\label{Boggs}
\end{align}
with $T_{q}=\hbar^2 q^2/(2m)$ and $\omega_{\bm{q}}=\sqrt{T_{q}^2+2n V(\bm{q})T_{q}}$ being the free particle and Bogoliubov dispersion, respectively, and $n=N/V$ is the boson density.

The condensate depletion, which is supposed to be small, is equal to
\begin{align}
\frac{n-n_0}{n}&=\frac{1}{n}\int\frac{\mathrm{d}^3q}{(2\pi)^3}n_{\bm{q}}, \label{conddep}\\
n_{\bm{q}}&=\frac{1}{2}\left(\frac{T_{q}+n V(\bm{q})}{\omega_{\bm{q}}}-1\right), \label{ocnum}
\end{align}
where $n_{\bm{q}}= \langle\hat{a}^{\dag}_{\bm{q}}\hat{a}_{\bm{q}}\rangle$ denotes the average occupation numbers of the bosons for $\bm{q}\not=0$, and $n_0=N_0/V$ is the density of the Bose-Einstein condensate.

An effective pseudopotential for the dipolar gas was suggested in Ref.~\cite{Yi00}:
\begin{align}\label{Vpseudo}
V(\bm{r})=\frac{4\pi\hbar^2 a}{m}\delta(\bm{r})-\frac{2d^2}{r^3}P_2(\bm{e}_{d}\cdot\bm{e}_{r}).
\end{align}
Here $P_2(x)=(3x^2-1)/2$ is the Legendre polynomial of the second order, $d$ is the absolute value of the dipole, and $\bm{e}_{r}$ is the unit vectors along the directions of the relative coordinate. The dipoles are supposed to be aligned along the same direction $\bm{e}_{d}$  by a homogeneous external field. The first term is the zero-range interaction with the scattering length $a$, which is assumed to be positive for the stability of the system (see the detailed analysis in Sec.~\ref{sec:thermstab} below). The Fourier transform of the pseudopotential (\ref{Vpseudo}) gives us the \emph{low-momentum} scattering amplitude \cite{Schutzhold06} for $q\not=0$,
\begin{align}\label{uqeffps}
U(\bm{q})=\frac{4\pi\hbar^2 a}{m} +\frac{8\pi d^2}{3}P_2(\bm{e}_{d}\cdot\bm{e}_{q})
\end{align}
with $\bm{e}_{q}$ being the unit vector along the momentum $\bm{q}$. Below in Sec.~\ref{sec:lm_scat}, we  give a simple derivation of this equation from the two-body scattering problem.

One can see from Eq.~(\ref{uqeffps}) that due to the long-range nature of the dipolar forces, the value of $V(\bm{q})=U(\bm{q})$ at $q=0$ is not defined. This is because the Bogoliubov model is applicable for the weak short-range potentials. In three dimensions, they should decay at large distance as $1/r^\alpha$ with $\alpha>3$ or faster. Then the ground-state energy \emph{cannot} be calculated by using directly the pseudopotential (\ref{Vpseudo}).

Nevertheless, it is possible to consider the dipole-dipole interaction as a limiting case of a short-range interaction with respect to some parameter. We shall call this procedure regularization and take a look at it in the next subsection.

\subsection{The dipole-dipole interaction and its regularization}
\label{sec:ddint}

The dipole-dipole interaction is given by (see, e.g., Ref.~\cite{Jackson62:book})
\begin{align}
 V_{\mathrm{dd}}(\bm{r})=\frac{\bm{d}_{1}\cdot\bm{d}_{2}\,r^2-3(\bm{d}_{1}\cdot\bm{r}) (\bm{d}_{2}\cdot \bm{r})}{r^5},  \label{ddint}
\end{align}
where $\bm{d}_{1}$ and $\bm{d}_{2}$ are the vectors of two dipoles. It is an even function of the relative coordinate $\bm{r}$. In order to calculate its Fourier transform, one can use \cite{Eberlein05} the identities
\begin{align}
\delta_{ij}\frac{1}{r^3}-\frac{3x_i x_j}{r^5}&=-\frac{\partial^2}{\partial x_i \partial x_j} \frac{1}{r}-\frac{4\pi}{3}\delta_{ij}\delta(\bm{r}),\label{identity}\\
\frac{1}{r}&=\int\frac{\mathrm{d}^3q}{(2\pi)^3}\frac{4\pi}{q^2}e^{i\bm{q}\cdot\bm{r}},\label{CoulombFT}
\end{align}
where $\delta_{ij}$ and $\delta(\bm{r})$ are the Kronecker delta and the Dirac $\delta$-function, respectively, and $x_i$ with $i=1,2,3$ are the Cartesian coordinates. Then the Fourier transform amounts to
\begin{align}\label{ddFT}
V_{\mathrm{dd}}(\bm{q})=\frac{4\pi(\bm{d}_{1}\cdot\bm{q})(\bm{d}_{2}\cdot\bm{q})}{q^2} -\frac{4\pi}{3}(\bm{d}_{1}\cdot\bm{d}_{2}).
\end{align}

As one can see, the Fourier transform (\ref{ddFT}) depends on the direction of the momentum $\bm{q}$ but not its absolute value. For this reason, the limit $q\to0$ does not exist. This is a direct consequence of the long-range nature of the dipole-dipole interaction.

As is mentioned above, it is convenient to regularize the long-range interaction (\ref{ddint}) by presenting it as a limiting case of a short-range interaction with respect to some parameter. Equations (\ref{identity}) and (\ref{CoulombFT}) suggest using the screened Coulomb potential
\begin{align}
V_{\varkappa}(\bm{r})=-(\bm{d}_{1}\cdot\nabla)(\bm{d}_{2}\cdot\nabla)\frac{e^{-\varkappa r}}{r} -\frac{4\pi}{3}(\bm{d}_{1}\cdot\bm{d}_{2})\delta(\bm{r})
\label{ddintscr}
\end{align}
in the limit of infinite range of the screening $\varkappa\to0$. We do not need to write down explicitly the derivatives of the function ${e^{-\varkappa r}}/{r}$, it is sufficient to use the expression $4\pi/(q^2+\varkappa^2)$ for its Fourier transform, which leads to
\begin{align}\label{ddscrFT}
V_{\varkappa}(\bm{q})=\frac{4\pi(\bm{d}_{1}\cdot\bm{q})(\bm{d}_{2}\cdot\bm{q})}{q^2+\varkappa^2} -\frac{4\pi}{3}(\bm{d}_{1}\cdot\bm{d}_{2}).
\end{align}
Now this equation is well defined at $q=0$ for arbitrary $\varkappa\not=0$ and equals $-{4\pi}(\bm{d}_{1}\cdot\bm{d}_{2})/{3}$, while for nonzero momentum, the limit $\varkappa\to0$ reproduces  Eq.~(\ref{ddFT}). If the dipoles are aligned along the same direction $\bm{d}_{1}= \bm{d}_{2}= d\bm{e}_{d}$, one can write finally the regularized dipole-dipole interaction as
\begin{align}\label{ddfinFT}
V_{\mathrm{dd}}(\bm{q})=\frac{4\pi d^2}{3}\times\begin{cases}
                          2P_2(\bm{e}_{d}\cdot\bm{e}_{q}), & \text{for } q\not=0,\\
                          (-1), & \text{for } q=0.
                        \end{cases}
\end{align}
The discontinuity at $q=0$ is typical for effective long-range potentials (see, e.g., Ref.~\cite{Foldy61}).

It should be emphasized that not every regularization leads to the same behavior of the Fourier transform of the dipole-dipole interaction. For instance, if we apply the regularization with the same exponentially dependent prefactor $\exp(-\varkappa\, r)$ to the dipolar potential itself (see detailed discussions of this kind of regularizations in  Ref.~\cite{Yukalov16}), we find that its Fourier component at $q=0$ is equal to zero in the limit $\varkappa\to 0$. The choice of the regularization (\ref{ddintscr}) can be justified by the self-consistency of the results obtained in this way. The mean-field term in the expansion of the energy, first, is compatible with the thermodynamic  stability of the system, and, second, admits a clear physical interpretation (see Secs.~\ref{sec:thermstab} and \ref{sec:interp}, respectively, below).

\subsection{The low-density expansion of the ground-state energy, chemical potential, and condensate depletion}
\label{grstdepl}

Following the scheme intended in Sec.~\ref{sec:bog}, the contribution of the dipole part in the scattering amplitude (\ref{uqeffps}) should be regularized in accordance with Eq.~(\ref{ddfinFT}):
\begin{align}\label{uqeffBog}
U_{\mathrm{eff}}(\bm{q})=\frac{4\pi\hbar^2 a}{m}\times\begin{cases}
[1+2\epsilon_{\mathrm{dd}} P_2(\bm{e}_{d}\cdot\bm{e}_{q})], & \text{for } q\not=0,\\
(1-\epsilon_{\mathrm{dd}}), & \text{for } q=0,
                        \end{cases}
\end{align}
which makes it applicable to the Bogoliubov model. We introduce the standard notation \cite{Pitaevskii16:book} for the ratio $\epsilon_{\mathrm{dd}}=r_{\mathrm{dd}}/a$, where $r_{\mathrm{dd}}=d^2m/(3\hbar^2)$ is the effective dipole range.

Replacing $V(\bm{q})$ by $U_{\mathrm{eff}}(\bm{q})$ in Eqs.~(\ref{Boggs}) and (\ref{conddep}) yields
\begin{align}
  &\varepsilon = \frac{2\pi\hbar^2}{m}an\left[1-\epsilon_{\mathrm{dd}}+\frac{128}{15\sqrt{\pi}}\sqrt{n a^3}{\cal Q}_{5}(\epsilon_{\mathrm{dd}})\right], \label{expen}\\
  &\frac{n-n_0}{n}= \frac{8}{3\sqrt{\pi}}\sqrt{n a^3}{\cal Q}_{3}(\epsilon_{\mathrm{dd}}).  \label{expdepl}
\end{align}
Here we use the notation of Ref.~\cite{Lima12}
\begin{align}
{\cal Q}_{n}(y)=(1-y)^{n/2}{}_{2}F_{1}\left(\frac{1}{2},-\frac{n}{2};\frac{3}{2};-\frac{3y}{1-y}\right)
\label{Qn}
\end{align}
with ${}_{2}F_{1}$ being the hypergeometric function \cite{abr_steg64}. The Taylor expansion of the function ${\cal Q}_{n}(y)$ in $y$ near zero takes the form ${\cal Q}_{n}(y)= 1+y(n^2-2n)/10 +\cdots$. The function ${\cal Q}_{n}$ is a polynomial at even $n$, and it can be written through elementary functions for odd $n$. In the particular cases $n=3$ and $n=5$, it becomes
\begin{align}
{\cal Q}_{3}(y)=&\frac{5+y}{8}\sqrt{1+2y}+\frac{\sqrt{3}}{8}\frac{(y-1)^2}{\sqrt{y}}p(y), \label{q3an}\\
{\cal Q}_{5}(y)=&\frac{33+12y+27y^2}{48}\sqrt{1+2y}-\frac{5\sqrt{3}}{48}\frac{(y-1)^3}{\sqrt{y}}p(y),\label{q5an}\\
p(y)=&\ln(\sqrt{3y}+\sqrt{1+2y})-\frac{1}{4}\ln[(1-y)^2]. \nonumber
\end{align}
Being written in this form, the functions ${\cal Q}_{3}(y)$ and ${\cal Q}_{5}(y)$ are formally defined for an arbitrary non-negative argument.

Note that the first integral in the r.h.s. of Eq.~(\ref{Boggs}) diverges as $\int\mathrm{d}^3q/{q^{2}}$, and the divergent integral should merely be omitted \cite{llvol9_81}. The nature of this divergence is discussed in Sec.~\ref{sec:universality} below.

The term in the expansion of the energy (\ref{expen}) proportional to the density is the Hartree mean-field energy, while the term  proportional to $n^{3/2}$ is the Lee-Huang-Yang correction. It is associated with the zero-point energy of the Bogoliubov quasiparticles, given by the term $\omega(\bm{q})/2$ in the second integral in the r.h.s. of Eq.~(\ref{Boggs}).

The chemical potential
\begin{align}
\mu=\frac{4\pi\hbar^2}{m}an\left[1-\epsilon_{\mathrm{dd}}+\frac{32}{3\sqrt{\pi}}\sqrt{n a^3}{\cal Q}_{5}(\epsilon_{\mathrm{dd}})\right] \label{mudipolar}
\end{align}
is easily obtained from the ground-state energy per particle with the thermodynamic relation $\mu=\partial (n \varepsilon)/\partial n$.

The expansion parameter ${na^3}$, which is called the gas parameter, is supposed to be small. In the absence of the dipolar forces ($\epsilon_{\mathrm{dd}}=0$), the expansion for the condensate depletion (\ref{expdepl}) reproduces the Bogoliubov result \cite{Bogoliubov47}, and the expansions for the energy (\ref{expen}) and chemical potential (\ref{mudipolar}) coincide with that of Lee, Huang, and Yang \cite{Lee57}.

Let us discuss how the jump in the effective scattering amplitude (\ref{uqeffBog}) at $q=0$ affects the interaction term in the Gross-Pitaevskii energy functional. The interaction term is obtained by integration $E_{\mathrm{int}}=\frac{1}{2}\int\mathrm{d}^3r\mathrm{d}^3r' V_{\mathrm{eff}}(\bm{r}-\bm{r}')|\Phi(\bm{r})|^2|\Phi(\bm{r}')|^2$ with $V_{\mathrm{eff}}(\bm{r})$ being the Fourier transform of the effective scattering amplitude (\ref{uqeffBog}). Since the jump takes place at only one point in the momentum space, it is negligible for almost all inhomogeneous configurations of the system, and the Gross-Pitaevskii interaction term has the standard form $E_{\mathrm{int}} =\frac{1}{2}\int\mathrm{d}^3r\left[g|\Phi(\bm{r})|^4 +|\Phi(\bm{r})|^2|\int\mathrm{d}^3r' V_{\mathrm{dd}}(\bm{r}-\bm{r}')|\Phi(\bm{r}')|^2\right]$, where  $V_{\mathrm{dd}}(\bm{r})$ is the dipole-dipole interaction given by Eq.~(\ref{ddint}) when the dipoles are parallel. However, when the condensate wave function $\Phi(\bm{r})$ is a constant, as for the homogeneous dipolar gas, the contribution of the zero Fourier component becomes decisive.

\subsection{Thermodynamic stability of the homogeneous dipolar gas and conditions for droplets}
\label{sec:thermstab}

The stability of the ground state is determined by the thermodynamic relation
$(\partial \mu/\partial n)_{T}>0$. We obtain at zero temperature from Eq.~(\ref{mudipolar})
\begin{align}\label{TDst}
\frac{\partial \mu}{\partial n}=\frac{4\pi\hbar^2}{m}a\left[1-\epsilon_{\mathrm{dd}}+\frac{16}{\sqrt{\pi}}\sqrt{n a^3}{\cal Q}_{5}(\epsilon_{\mathrm{dd}})\right]>0.
\end{align}
If $a > r_{\mathrm{dd}}$ then the dipolar gas is stable. This matches well with the analysis of the Bogoliubov spectrum
\begin{align}
\label{bogspectr}
\omega_{\bm{q}}=\sqrt{T_{q}^2+2gnT_{q}[1+2\epsilon_{\mathrm{dd}} P_2(\bm{e}_{d}\cdot\bm{e}_{q})]},
\end{align}
where $g={4\pi\hbar^2a}/{m}$ is the standard coupling parameter. If the condition $a > r_{\mathrm{dd}}$ is satisfied then the spectrum is real and positive for arbitrary $\bm{q}$; otherwise the expression under the square root becomes negative at the minimum when $\bm{e}_{d}\cdot\bm{e}_{q}=0$.

However, the presence of the Lee-Huang-Yang correction in Eq.~(\ref{TDst}) allows the gas to be stable even at $\epsilon_{\mathrm{dd}}$ greater than one. Indeed, ${\cal Q}_{5}(\epsilon_{\mathrm{dd}})= {3 \sqrt{3}}/{2}+{15} \sqrt{3} (\epsilon_{\mathrm{dd}} -1)/{8}+\cdots$ in the vicinity of $\epsilon_{\mathrm{dd}}=1$, and the condition (\ref{TDst}) tells us that for $n_{\mathrm{cr}} <n\ll 1/a^3$ the dipolar gas can be stable, where the critical density is given by
\begin{align}\label{ncr}
n_{\mathrm{cr}}=\frac{\pi(\epsilon_{\mathrm{dd}}-1)^2}{1728 a^3\left[1+\frac{5}{4}(\epsilon_{\mathrm{dd}}-1)\right]^2}.
\end{align}
Even if the density is less than critical, the system cannot completely collapse, because it can be fragmented into small sufficiently dense subsystems called droplets. Then the stability is reached at the expense of inhomogeneity.

Earlier the droplets were predicted in a condensed Bose-Bose mixture \cite{Petrov15}, and they have been observed experimentally in dipolar gases \cite{pfau16,pfau16n,Chomaz16}.

Note that for $\epsilon_{\mathrm{dd}}>1$ the spectrum can take complex values, which implies that the stability appears to be impossible in the
homogeneous regime. This question is beyond the scope of this paper; we simply point out that the inhomogeneous geometry can change the spectrum accordingly without significant renormalization of the Lee-Huang-Yang correction.

\subsection{The physical interpretation of the dipole contribution into the leading term of the ground-state energy}
\label{sec:interp}

The term $-2\pi\hbar^2a\epsilon_{\mathrm{dd}}n/m =-2\pi d^2n/3$ in the expansion (\ref{expen}) for the energy appears due to the regularization discussed in Sec.~\ref{sec:ddint}. A simple physical interpretation can be given for this term. Let us consider the Bose gas of  electrical dipoles aligned with an electric field. In the field $\bm{E}$, a dipole $\bm{d}$ has the energy $\Delta \varepsilon =-(\bm{E}\cdot\bm{d})$. In a dielectric medium, the local electric field that a dipole ``feels'' is not just an average macroscopic field in the medium. It contains a local correction called Lorentz correction, which arises due to the distant dipoles: $\bm{E}_{\mathrm{lf}}=\bm{E}_{\mathrm{ext}}+4\pi\bm{P}/3$ (see, e.g., Ref.~\cite{Jackson62:book}). Here $\bm{P}=n\bm{d}$ is the polarizability of the medium, that is, the total dipole moment per unit volume, and $\bm{E}_{\mathrm{ext}}$ is the external field. We obtain for the energy of the dipole
\begin{align}\label{endipole}
\Delta \varepsilon=-E_{\mathrm{ext}}d-4\pi d^2n/3,
\end{align}
because $\bm{d}$ and $\bm{E}_{\mathrm{ext}}$ are parallel. The first term in the r.h.s. of Eq.~(\ref{endipole}) is independent of the density and thus gives the constant shift of the energy per particle, which can be omitted.

Thus the energy $-4\pi d^2n/3$ is nothing else but the classical electrostatic energy of a dipole in the field created by the other dipoles. When calculating the energy per particle, the factor one-half is needed to compensate the double counting of the interactions between all pairs of the dipoles. The same interpretation is valid for the magnetic dipoles.

\subsection{The long-range asymptotics of the normal and anomalous one-particle correlation functions}
\label{sec:corr_as}

By definition, the normal and anomalous one-particle correlation functions in the Bogoliubov model are equal to $\langle\hat{\psi}^{\dag}(\bm{r}_1) \hat{\psi}(\bm{r}_2)\rangle$ and $\langle\hat{\psi}(\bm{r}_1) \hat{\psi}(\bm{r}_2)\rangle$, respectively. Here $\hat{\psi}$ and $\hat{\psi}^{\dag}$ are the Bose field operators, whose Fourier transforms $\hat{a}_{\bm{q}}$ and $\hat{a}^{\dag}_{\bm{q}}$ are the annihilation and creation boson operators of a particle with momentum $\bm{q}$, respectively. Following Bogoliubov, one can separate the contribution of the condensate operators, replacing them by the $C$-numbers \cite{U1sym}:
$\hat{\psi}(\bm{r}) =\sqrt{n_{0}} +\sum_{\bm{q}\not=0}\hat{a}_{\bm{q}}e^{i\bm{q}\cdot\bm{r}}/\sqrt{V}$ and $\hat{\psi}^{\dag}(\bm{r}) =\sqrt{n_{0}} +\sum_{\bm{q}\not=0}\hat{a}^{\dag}_{\bm{q}}e^{-i\bm{q}\cdot\bm{r}}/\sqrt{V}$. In the thermodynamic limit we obtain
\begin{align}
\langle\hat{\psi}^{\dag}(\bm{r}_1)\hat{\psi}(\bm{r}_2)\rangle&=n_{0}+\int\frac{\mathrm{d}^3q}{(2\pi)^3} \langle\hat{a}^{\dag}_{\bm{q}}\hat{a}_{\bm{q}}\rangle e^{-i \bm{q}\cdot(\bm{r}_1-\bm{r}_2)}, \label{normcorr}\\
\langle\hat{\psi}(\bm{r}_1)\hat{\psi}(\bm{r}_2)\rangle&=n_{0}+\int\frac{\mathrm{d}^3q}{(2\pi)^3}\langle\hat{a}_{\bm{q}}\hat{a}_{-\bm{q}}\rangle e^{i \bm{q}\cdot(\bm{r}_1-\bm{r}_2)}, \label{anormcorr}
\end{align}
which depend only on the relative distance $\bm{r}_1-\bm{r}_2$ by virtue of the translational invariance. The normal correlator $\langle\hat{a}^{\dag}_{\bm{q}}\hat{a}_{\bm{q}}\rangle$ is the average occupation number (\ref{ocnum}), while the anomalous correlator
is given by \cite{Bogoliubov47}
\begin{align}
\langle\hat{a}_{\bm{q}}\hat{a}_{-\bm{q}}\rangle=-\frac{1}{2}\frac{n U_{\mathrm{eff}}(\bm{q})}{\omega_{\bm{q}}} \label{anomav}
\end{align}
with the Bogoliubov spectrum (\ref{bogspectr}). The anomalous correlator (\ref{anormcorr}) can be interpreted as the two-body wave function in the Bose-Einstein condensate \cite{cherny98,cherny00}.

\begin{figure}[!htbp]
\centerline{\includegraphics[width=.75\columnwidth]{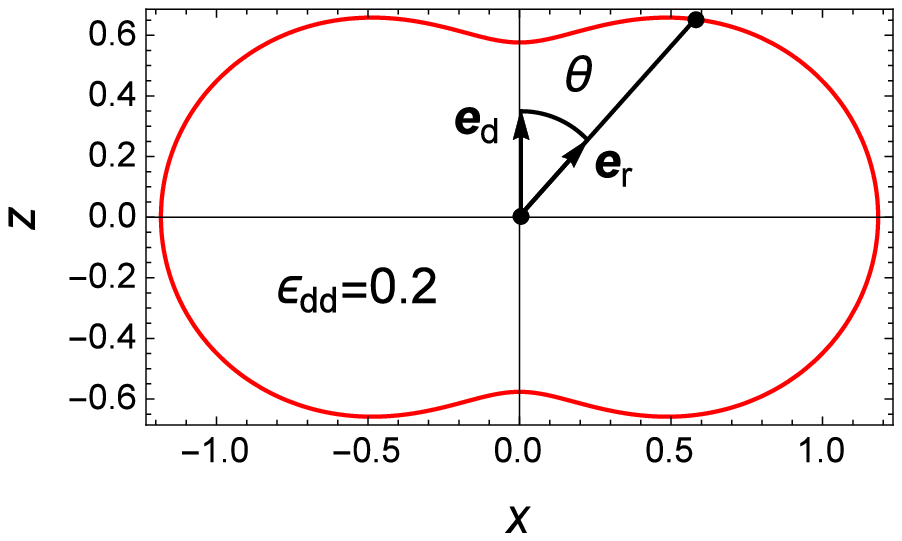}}
\centerline{\includegraphics[width=.75\columnwidth]{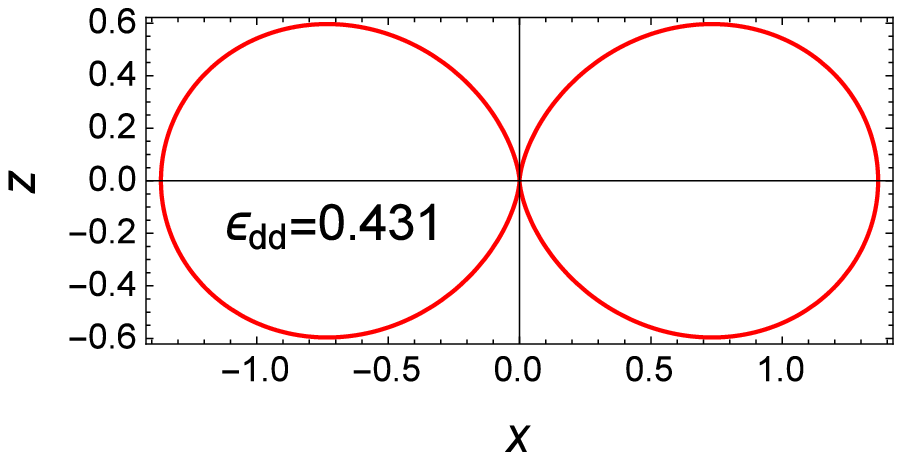}}
\centerline{\includegraphics[width=.75\columnwidth]{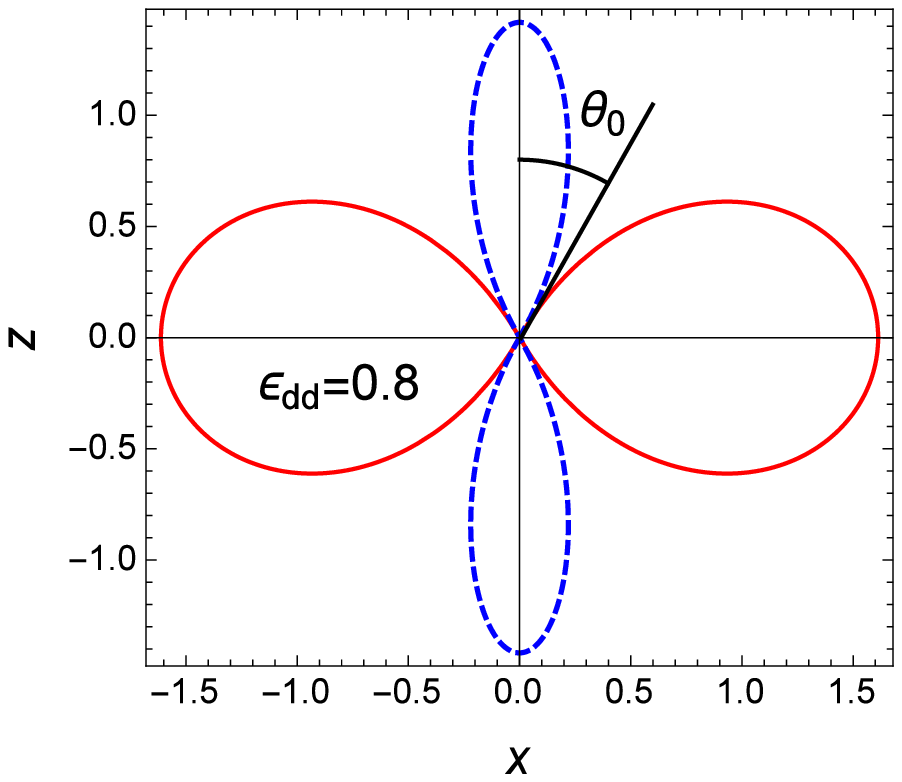}}
\caption{\label{fig:fu} Polar plot of the anisotropic prefactor $f(\epsilon_{\mathrm{dd}},\theta)$ in the asymptotics of the one-particle correlation functions (\ref{normas}) and (\ref{anomas}) at different values of the dipolar ratio $\epsilon_{\mathrm{dd}}$ (that is, $f$ is shown as a distance between the origin and a point on the plot as a function of the angle; see the upper panel). The prefactor is given by Eq.~(\ref{fueq}) with $(\bm{e}_{d}\cdot\bm{e}_{r})=\cos\theta$ and $\bm{e}_{d}$ being parallel to the $z$-axis. When the ratio exceeds the critical value $\epsilon_{\mathrm{dd}\,\mathrm{c}}=0.431\ldots$ (the middle panel), the prefactor becomes negative for $\theta<\theta_{\mathrm{0}}$ and $\theta>\frac{\pi}{2}-\theta_{\mathrm{0}}$ [here the angle $\theta_{\mathrm{0}}$ is given by Eq.~(\ref{thcr})]. The positive and negative values of the prefactor are shown in solid (red) and dashed (blue) lines, respectively.
 }
 \end{figure}
The long-range asymptotics of the correlators in real space are determined by their Fourier transforms in the vicinity of $\bm{q}=0$~\cite{erdelyi56:book}. When $q\to0$, we find from Eqs.~(\ref{ocnum}), (\ref{bogspectr}) and (\ref{anomav})
\begin{align}\label{nqpsiqsmall}
\langle\hat{a}^{\dag}_{\bm{q}}\hat{a}_{\bm{q}}\rangle
\simeq -\langle\hat{a}_{\bm{q}}\hat{a}_{-\bm{q}}\rangle
\simeq \frac{\sqrt{\pi n a}}{q}\sqrt{1+2\epsilon_{\mathrm{dd}} P_2(\bm{e}_{d}\cdot\bm{e}_{q})}.
\end{align}
The singularity $1/q$ for $q\to 0$ appears in accordance with the Bogoliubov theorem \cite{bogoliubov61:book}. From the physical point of view, the momenta are small as long as the Bogoliubov spectrum is linear in $q$, which takes place when $q\ll1/\xi$. Here
\begin{align}\label{xi}
 \xi= \frac{1}{2\sqrt{\pi a n(1-\epsilon_{\mathrm{dd}})}}
\end{align}
is the healing length.

Substituting Eq.~(\ref{nqpsiqsmall}) into Eqs.~(\ref{normcorr}) and (\ref{anormcorr}) yields \cite{divergence} the asymptotics of the correlators for $r\gg \xi$
\begin{align}
\frac{\langle\hat{\psi}^{\dag}(\bm{r})\hat{\psi}(0)\rangle}{n} &\simeq 1+\frac{1}{2r^2}\sqrt{\frac{an}{\pi^3}}\,f(\epsilon_{\mathrm{dd}},\theta), \label{normas}\\
\frac{\langle\hat{\psi}(\bm{r})\hat{\psi}(0)\rangle}{n}&\simeq 1 - \frac{1}{2r^2}\sqrt{\frac{an}{\pi^3}}\,f(\epsilon_{\mathrm{dd}},\theta), \label{anomas}
\end{align}
where the anisotropic factor is given by
\begin{align}
f&(\epsilon_{\mathrm{dd}},\theta)= \sqrt{1+2\epsilon_{\mathrm{dd}}}\left(1+\frac{u}{2}\ln\frac{1-u}{1+u}\right), \label{fueq}\\ u&=\cos\theta\sqrt{\frac{3\epsilon_{\mathrm{dd}}}{1+2\epsilon_{\mathrm{dd}}}},\quad \cos\theta=(\bm{e}_{d}\cdot\bm{e}_{r}).\nonumber
\end{align}
Here we assume that the dipoles are parallel to the $z$-axis and $\theta$ is a polar angle in the spherical coordinate system.
The factor can be expanded into the Legendre polynomials of the angle variable
\begin{align}
f(u)=&-\sqrt{\frac{1\!-\!\epsilon_{\mathrm{dd}}}{1\!+\!2\epsilon_{\mathrm{dd}}}}\sum_{l=0}^{\infty}P_{2l}(\bm{e}_{d}\cdot\bm{e}_{r})
\frac{(l\!+\!\frac{1}{4})\Gamma(l\!+\!1)\Gamma(l\!-\!\frac{1}{2})}{\Gamma(2l\!+\!\frac{3}{2})} \nonumber\\
&\times{}_{2}F_{1}\left(l-\frac{1}{2},l+\frac{1}{2};2l+\frac{3}{2};-z\right)z^l,
\label{fuLeg}
\end{align}
where $z={3\epsilon_{\mathrm{dd}}}/({1-\epsilon_{\mathrm{dd}}})$ and $\Gamma(x)$ is the gamma function.

The obtained asymptotics for $r\gg \xi$ are universal, that is, they depend only on the scattering length and dipolar range.  By contrast, the behaviour of the correlators $\langle\hat{\psi}^{\dag}(\bm{r}) \hat{\psi}(0)\rangle$ and $\langle\hat{\psi}(\bm{r})\hat{\psi}(0)\rangle$ is not universal when $r\ll \xi$ \cite{cherny00}.

At $\epsilon_{\mathrm{dd}}=0$, the contribution of the dipolar potential is zero, and the asymptotics of the correlators (\ref{normas}) and (\ref{anomas}) for $r\gg\xi$ coincides with that for radially symmetric short-range potentials \cite{cherny01}. As the ratio $\epsilon_{\mathrm{dd}}$ grows, the asymptotics strongly depends on the angle $\theta$ between $\bm{r}$ and $\bm{d}$ through prefactor $f(u)$, see Fig.~\ref{fig:fu}. Starting from the critical value $\epsilon_{\mathrm{dd}\,\mathrm{c}}=0.431\ldots$, the sign of the prefactor changes at the angles
\begin{align}\label{thcr}
 \theta_{\mathrm{0}}(\epsilon_{\mathrm{dd}})=\arccos\sqrt{u_{\mathrm{c}}\frac{1+2\epsilon_{\mathrm{dd}}}{3\epsilon_{\mathrm{dd}}}}
\end{align}
and $\frac{\pi}{2}-\theta_{\mathrm{0}}$, where $u_{\mathrm{c}}=0.833\ldots$ is the root of the equation $1+\frac{u}{2}\ln\frac{1-u}{1+u}=0$.

In principle, the normal correlator should manifest itself in interference patterns of two small clouds of atoms, which are ejected from different parts of the sample. Certainly, an observation of the decaying part of the correlator might be a very difficult problem in practice.

\subsection{The long-range asymptotics of the pair distribution function}
\label{sec:gr_as}

\begin{figure}[!htbp]
\begin{center}
\includegraphics[width=.49\columnwidth]{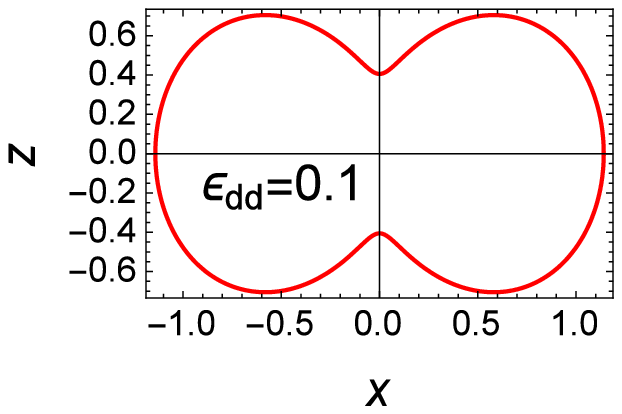}\includegraphics[width=.49\columnwidth]{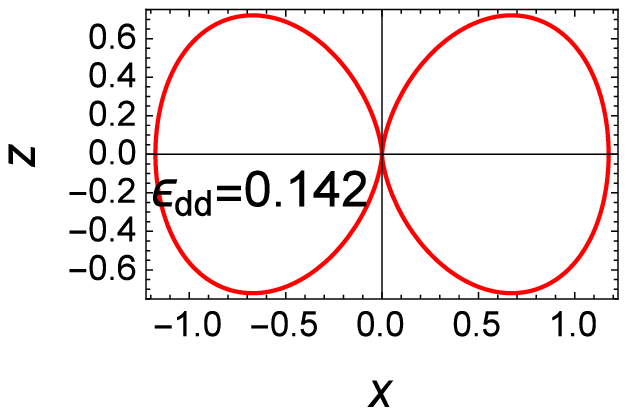}\\
\includegraphics[width=.37\columnwidth]{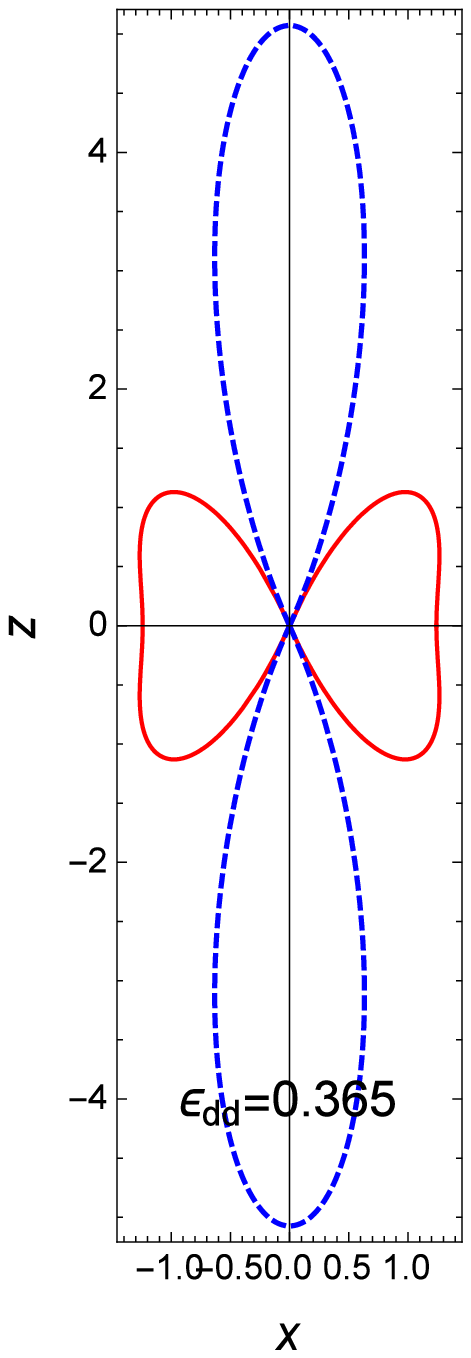}\hfil\includegraphics[width=.325\columnwidth]{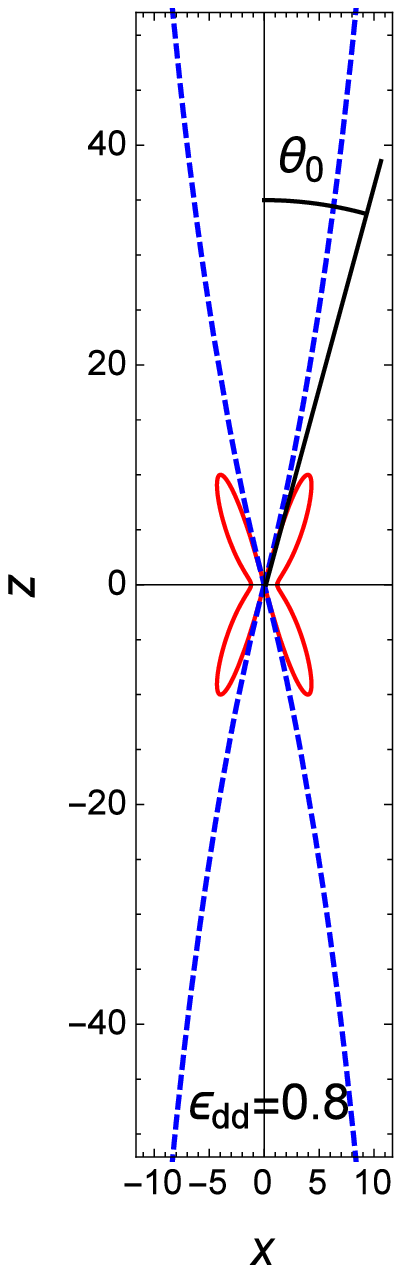}
\end{center}
\caption{\label{fig:hu} Polar plot of the anisotropic prefactor $h(\epsilon_{\mathrm{dd}},\theta)$ in the asymptotics of the pair distribution function (\ref{gras}) at different values of the dipolar ratio $\epsilon_{\mathrm{dd}}$. The notations are the same as in Fig.~\ref{fig:fu}. When the dipolar ratio exceeds the critical value $1/7$, the prefactor becomes negative within the two cones, whose axes coincide with the $z$ axis, the common cone vertex lies in the origin, and the apex angle $\theta_{\mathrm{0}}$ is given by Eq.~(\ref{th0dd}).
}
\end{figure}

By definition, the pair distribution function is proportional to the density-density correlator
\begin{align}\label{grdef}
g(\bm{r}_{1}-\bm{r}_{2})=\frac{\langle\hat{\rho}(\bm{r}_{1})\hat{\rho}(\bm{r}_{2})\rangle}{n^2},
\end{align}
where the density operator is equal to $\hat{\rho}(\bm{r})=\hat{\psi}^{\dag}(\bm{r})\hat{\psi}(\bm{r})$. Then the quantity $g(\bm{r}_{1}-\bm{r}_{2})/V$ is nothing else but the density of the conditional probability of finding one particle at the point $\bm{r}_{1}$ while another is at the point $\bm{r}_{2}$.

The pair distribution function is directly related to the static structure factor
\begin{align}\label{grsk}
g(\bm{r}) = 1+ \frac{1}{n}\int\frac{\mathrm{d}^3q}{(2\pi)^3}[S(\bm{q})-1]e^{i \bm{q}\cdot\bm{r}},
\end{align}
which is given by $S(\bm{q})=T_{q}/\omega_{\bm{q}}$ in the Bogoliubov theory. As in Sec.~\ref{sec:corr_as}, we can replace the structure factor by its low-momentum asymptotics, because we are looking for the long-range asymptotics of the pair distribution function. We have
\begin{align}\label{skas}
S(\bm{q})\simeq \frac{q}{4\sqrt{\pi n a}}\frac{1}{\sqrt{1+2\epsilon_{\mathrm{dd}} P_2(\bm{e}_{d}\cdot\bm{e}_{q})}}.
\end{align}
Substituting this equation into Eq.~(\ref{grsk}) yields after a little algebra \cite{divergence}
\begin{align}
g(\bm{r}) =& 1 -\frac{1}{r^4}\frac{1}{4\pi^{5/2}n^{3/2}a^{1/2}}h(\epsilon_{\mathrm{dd}},\theta), \label{gras}\\[.5mm]
h(\epsilon_{\mathrm{dd}},\theta)=&\sqrt{1\!+\!2\epsilon_{\mathrm{dd}}}
\frac{1\!+\!7{\epsilon_{\mathrm{dd}}}\!+\!10\epsilon_{\mathrm{dd}}^2\!-\!3{\epsilon_{\mathrm{dd}}}({\epsilon_{\mathrm{dd}}}\!+\!5)\cos^2\theta} {\left[1+{\epsilon_{\mathrm{dd}}}(2-3\cos^2\theta)\right]^3},\label{heth}
\end{align}
where, as usual, $\cos\theta=(\bm{e}_{d}\cdot\bm{e}_{r})$. The asymptotics (\ref{gras}) is universal, and it is applicable when $r\gg\xi$.

In the absence of the dipolar forces ($\epsilon_{\mathrm{dd}}=0$), the prefactor $h(\epsilon_{\mathrm{dd}},\theta)$ is equal to one, and the second term in the r.h.s. of Eq.~(\ref{gras}) reproduces the well-known decay $1/r^4$ for a radially symmetric short-range potential \cite{Isihara64}. Note that the $1/r^4$ decay of $g(\bm{r})-1$ arises due to the \emph{linear} dependence of the static structure factor at small momentum. In spite of the long-range nature of the dipolar forces, the dependence is still linear, although anisotropic through the angular dependent coefficient of $q$ [see Eq.~(\ref{skas})].

The prefactor $h(\epsilon_{\mathrm{dd}},\theta)$ becomes strongly anisotropic even at quite small values of the dipolar ratio $\epsilon_{\mathrm{dd}}$, and changes sign when $\epsilon_{\mathrm{dd}}>1/7$ at the angle
\begin{align}\label{th0dd}
\theta_{\mathrm{0}}(\epsilon_{\mathrm{dd}})=\arccos\sqrt{\frac{1+7 \epsilon_{\mathrm{dd}} +10 \epsilon_{\mathrm{dd}}^2}{3 \epsilon_{\mathrm{dd}}  (\epsilon_{\mathrm{dd}} +5)}}
\end{align}
and $\frac{\pi}{2}-\theta_{\mathrm{0}}$, see Fig.~\ref{fig:hu}. The angle $\theta_{\mathrm{0}}$ determines the solid angle in real space where the prefactor $h(\epsilon_{\mathrm{dd}},\theta)$ is negative: $\theta<\theta_{\mathrm{0}}$ and $\theta>\frac{\pi}{2}-\theta_{\mathrm{0}}$. The angle $\theta_{\mathrm{0}}(\epsilon_{\mathrm{dd}})$ reaches maximum at $\epsilon_{\mathrm{dd}}=\frac{1+6\sqrt{6}}{43}=0.365\ldots$ and tends to zero as $\theta_{\mathrm{0}}(\epsilon_{\mathrm{dd}})\simeq\sqrt{(1-\epsilon_{\mathrm{dd}})/3}$ for $\epsilon_{\mathrm{dd}}\to 1$.

The absolute value of the prefactor grows fast for $\epsilon_{\mathrm{dd}}\to 1$. In order to estimate how the growth influences the asymptotics, we rewrite Eq.~(\ref{gras}) as
\begin{align}\label{grasxi}
 g(\bm{r}) = 1 -\frac{1}{(r/\xi)^4}\frac{4\sqrt{n a^3}}{\sqrt{\pi}}(1-\epsilon_{\mathrm{dd}})^2 h(\epsilon_{\mathrm{dd}},\theta)
\end{align}
with $\xi$ being the healing length (\ref{xi}). When the vector $\bm{r}$ is parallel or antiparallel to the dipole direction and $\epsilon_{\mathrm{dd}}=1$, we find
\[
g(\bm{r})= 1 +\frac{1}{(r/\xi)^4}\frac{24\sqrt{n a^3}}{\sqrt{\pi}}.
\]
If the distance is of order of the healing length, we obtain the estimation $g(\bm{r})\simeq 1 +{24\sqrt{n a^3}}/{\sqrt{\pi}}$. Although the gas parameter $n a^3$ is supposed to be small, the positive correction to one might not be so small due to the prefactor $24/\sqrt{\pi}$. Thus, when the dipolar ratio is close to one, the pair distribution function, which gives the probability to find a particle in the vicinity of another particle, exceeds one for $r\simeq \xi$ and is located within the very narrow cones near $\theta=0$ and $\theta=\pi$, see Fig.~\ref{fig:hu}. This means that the dipolar gas should have some tendency to form filaments along the dipole direction when $\epsilon_{\mathrm{dd}}\simeq 1$.

This conclusion matches well the analysis of Ref.~\cite{Wachtler16} that shows a filament-like form of droplets in the regime $\epsilon_{\mathrm{dd}}>1$, which are markedly elongated along the $z$-axis.

\section{The scattering amplitude for realistic atomic interactions}
\label{sec:pseudo}

In this section, we consider the two-body scattering problem with zero relative momentum for the dipole-dipole interaction \emph{with the cutoff} at small distances. In particular, the scattering length is obtained as a perturbation series in a small parameter, the ratio of the dipolar range to the cutoff length. The results of Secs.~\ref{sec:lm_scat} and \ref{sec:Born} below are valid for the dipole-dipole interaction \emph{without} the long-range regularization, discussed above in Sec.~\ref{sec:ddint}.

\subsection{The low-momentum scattering amplitude and the asymptotics of the wave function}
\label{sec:lm_scat}

A realistic interaction in the dipolar Bose gas can be presented as the sum
\begin{align}\label{fullint}
V(\bm{r})=V_{0}(r)+V_{\mathrm{dd}}(\bm{r},r_0)
\end{align}
of a short-range potential $V_{0}(r)$, decreasing at large distances typically as $1/r^6$ \cite{Baranov12}, and the long-range dipole ``tail''
\begin{align}\label{ddintcut}
 V_{\mathrm{dd}}(\bm{r},r_0)=\begin{cases}
 d^2[{1-3(\bm{e}_{d}\cdot\bm{e}_{r})^2}]/{r^3} ,  &\text{for } r\geqslant r_0,\\
 0, &\text{for } r < r_0.
 \end{cases}
\end{align}
Here $r_0$ is a cutoff parameter of order of atomic size. The short-range potential is often modeled with a hard sphere of radius $r_0$ \cite{Ronen06}.

The Fourier transform of the long-range interaction (\ref{ddintcut}) is easily calculated as
\begin{align}\label{vddqr0}
V_{\mathrm{dd}}(\bm{q},r_0)=8\pi d^2 P_2(\bm{e}_{d}\cdot\bm{e}_{q}) \frac{j_{1}(r_0 q)}{r_0 q}.
\end{align}
Here $j_{1}(z)=(\sin z -z\cos z)/z^2$ is the spherical Bessel function of the first order \cite{abr_steg64}, whose Taylor series in the vicinity of $z=0$ is $j_{1}(z)=z/3+\cdots$. Therefore, the low-momentum asymptotics of the dipolar part of the interaction (\ref{ddintcut}) is independent of the cutoff parameter and coincides with the Fourier transform of the original dipole-dipole interaction given by Eq.~(\ref{ddFT}) with $\bm{d}_{1} =\bm{d}_{2}$.

For a radially symmetric short-range potential, the scattering of two particles with zero relative momentum is characterized by the scattering length $a$. The presence of the dipole long-range interaction strongly affects the scattering, and the low-momentum scattering amplitude for the interaction (\ref{fullint}) is given by Eq.~(\ref{uqeffps}) as discussed above in Sec.~\ref{sec:bog}. In accordance with the definition of the scattering amplitude, it is given by
\begin{align}\label{Uqdef}
U(\bm{q})=\int \mathrm{d}r^3\, e^{-i\bm{q}\cdot\bm{r}}V(\bm{r})\varphi(\bm{r}),
\end{align}
where the wave function $\varphi(\bm{r})$ is the solution of the Schr\"odinger equation $\nabla^2\varphi(\bm{r})=mV(\bm{r})\varphi(\bm{r})/\hbar^2$. The standard boundary conditions, imposed on it, are that the wave function is bounded at $r=0$ and tends to one when $r\to\infty$. Besides, the wave function should be symmetric under the exchange of the two bosons and hence an even function of $\bm{r}$.

The solution of the Schr\"odinger equation can be expanded into the spherical harmonics, which contains only components with even angular momentum $l$, since it should an even function of the radius vector. Besides, the total potential $V(\bm{r})$ is axially symmetric under rotations about the dipole direction $\bm{e}_{d}$, and, therefore, the $z$-component of the angular momentum $\hbar m$ is conserved. This means that for each given $m$, the Sch\"odinger equation is reduced to a chain of coupled differential equations, which relate the components with momenta $l-2$, $l$, and $l+2$. On the other hand, the boundary condition at $|\bm{r}|=r_0$ can be chosen radially symmetric, because for $r\leqslant r_0$, the solution is governed only by the symmetric potential $V_{0}(r)$. The asymptotics $\varphi(\bm{r})\simeq 1$ when $r\to \infty$ is radially symmetric as well. Then we can put all the components with $m\not=0$ and arbitrary $l$ equal to zero, because this solution obeys the Schr\"odinger equation with the imposed boundary conditions. Therefore, we are left only with the components with $m=0$, and the expansion takes the form
\begin{align}\label{expphi}
\varphi(\bm{r})=\sum_{l=0}^{\infty}\varphi_{2l}(r)P_{2l}(\bm{e}_{d}\cdot\bm{e}_{r}).
\end{align}
Here $P_{n}$ are the Legendre polynomials, and the components $\varphi_{n}$ are nothing else but the partial waves.

Let us derive the low-momentum asymptotics of the scattering amplitude from the boundary conditions. Substituting the full potential (\ref{fullint}) and expansion (\ref{expphi}) into Eq.~(\ref{Uqdef}) and taking the limit $q\to0$ yield precisely Eq.~(\ref{uqeffps}) with
\begin{align}
a= \frac{m}{\hbar^2}\int_{0}^{\infty}\!\! \mathrm{d}r\, r^2 V_{0}(r)\varphi_{0}(r)
                         -\frac{6r_{\mathrm{dd}}}{5}\int_{r_{0}}^{\infty} \mathrm{d}r \frac{\varphi_{2}(r)}{r}.
\label{aV0Vdd}
\end{align}
Here we assume that the ``scattering part'' $\varphi(\bm{r})-1$ of the wave function falls off quite rapidly, say, as $1/r^{\delta}$ with small positive  $\delta$ when $r\to\infty$ and use the orthogonality of Legendre polynomials \cite{abr_steg64} $\int_{-1}^{1}\mathrm{d}x P_m(x)P_k(x)=2\delta_{mk}/(2m+1)$. The anisotropic dipolar part of the scattering amplitude (\ref{uqeffps})
comes from the Fourier transform (\ref{vddqr0}) of the long-range dipole tail (\ref{ddintcut}).

Thus the cutoff does not influence the anisotropic part of the low-momentum asymptotics of the scattering amplitude, which results from the dipolar long-range tail only. By contrast, the dipolar contribution to the scattering length is quite sensitive to the cutoff parameter $r_0$. Moreover, the appropriate choice of $r_0$ even leads to a resonance \cite{Ronen06}.

The low-momentum behaviour of the scattering amplitude determines the long-range asymptotics of the wave function. Indeed, it follows from the Schr\"odinger equation that the Fourier transform of the scattering part $\varphi(\bm{r})-1$ of the wave function takes the form $-m U(\bm{q})/(\hbar^2q^2)$. This, together with Eq.~(\ref{uqeffps}), yields the asymptotics of the wave function at large distances
\begin{align}\label{phiasym}
\varphi(\bm{r})= 1- \frac{a-r_{\mathrm{dd}}P_2(\bm{e}_{d}\cdot\bm{e}_{r})}{r} +\cdots.
\end{align}
So, starting from the assumption that $\varphi(\bm{r})-1$ decreases quite rapidly at large distances, we are able to specify its main asymptotics up to a constant $a$. Certainly, the scattering length $a$ depends on $r_0$ and a shape of the short-range potential $V_0(r)$ and can be found only from the Schr\"odinger equation.

We emphasize that the wave function contains all partial waves with even momenta, because the Schr\"odinger equation relates any $\varphi_{2l}$ component to  $\varphi_{2l-2}$ and $\varphi_{2l+2}$. Nevertheless, the main asymptotic behavior of the wave function arises due to the two lowest components: $\varphi_{0}(r)\simeq1-a/r$ and $\varphi_{2}(r)\simeq r_{\mathrm{dd}}/r$ when $r\to\infty$. The components with higher angular momentum decay faster than $1/r$.

\subsection{The perturbation series for the dipolar potential with the cutoff}
\label{sec:Born}

If the scattering length and wave function in the absence of the dipolar forces are known and the value of the dipolar range is sufficiently small, the scattering length is obtained in the lowest approximation in $r_{\mathrm{dd}}$ by analogy with the Born perturbation series \cite{Migdal77:book}. In particular, for the hard sphere of radius $a_{\mathrm{sp}}$, we have $V_{0}(r)=+\infty$ for $r<a_{\mathrm{sp}}$ and zero elsewhere. The scattering length of the hard sphere coincides with its radius. When $a_{\mathrm{sp}}\leqslant r_{0}$, we obtain for $r_{\mathrm{dd}}\ll r_{0}$
\begin{align}\label{adBorn}
a=a_{\mathrm{sp}} -\frac{3}{25}\left( 2\frac{a^{2}_{\mathrm{sp}}}{r_0^{2}} - 7\frac{a_{\mathrm{sp}}}{r_0} +8\right)\frac{r^2_{\mathrm{dd}}}{r_0} + \cdots.
\end{align}
In the case $a_{\mathrm{sp}}=0$, this equation reproduces the second-order Born approximation for the scattering length (see Sec.~\ref{sec:Born_reg}). It should be emphasized that the perturbation expansion is possible only when the cutoff parameter $r_0$ is not equal to zero.

\begin{figure}[!htbp]
\centerline{\includegraphics[width=\columnwidth]{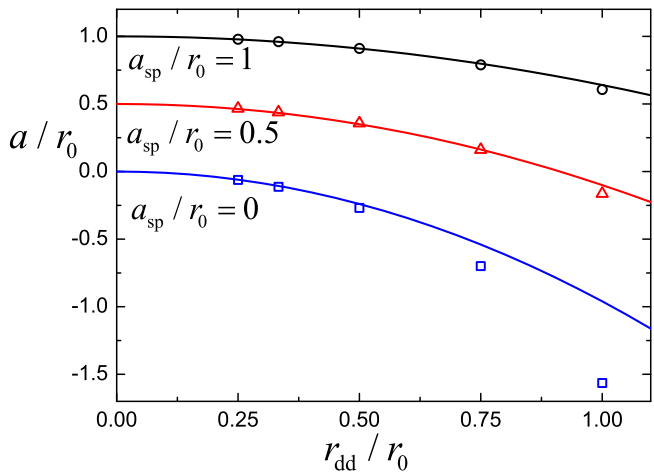}}
\caption{\label{fig:ardd} Scattering length $a$ vs the dipolar range $r_{\mathrm{dd}}$. The short-range potential is the hard sphere of radius $a_{\mathrm{sp}}$ and the dipole-dipole interaction is cut at small distances, see Eq.~(\ref{ddintcut}). All the parameters are shown in units of the cutoff parameter $r_{0}$. The numerical data are open squares, triangles, and circles for $a_{\mathrm{sp}}/r_{0}=0,0.5,1$, respectively. The data generated with the perturbative relation (\ref{adBorn}) are shown for the same values of $a_{\mathrm{sp}}$ in solid blue, red, and black lines, respectively. The smaller ratio $r_{\mathrm{dd}}/r_{0}$, the better it works. For $a_{\mathrm{sp}}=0$, the perturbative relation coincides with the second-order Born approximation.
 }
\end{figure}

This formula is checked numerically for a few values of the dipolar range; see Fig.~\ref{fig:ardd}. For the numerical solutions of the Schr\"odinger equation, we restrict ourselves to the zero and second components in the expansion over the Legendre polynomials: $\varphi(\bm{r})=\varphi_{0}(r)+\varphi_{2}(r)P_{2}(\bm{e}_{d}\cdot\bm{e}_{r})$.

\subsection{The scatting length for the regularized dipolar potential}
\label{sec:Born_reg}

The same regularization as in Sec.~\ref{sec:ddint} can be applied to the dipolar potential with the cutoff (\ref{ddintcut}): one should take the limit $\varkappa \to 0$ for the function defined as $V_{\varkappa}(\bm{r})=-(\bm{d}\cdot\nabla)^2({e^{-\varkappa r}}/{r})$ for $r\geqslant r_0$ and zero elsewhere. Then its Fourier transform for $q\to 0$ is still given by Eq.~(\ref{ddfinFT}). As a result, the new terms $-{4\pi\hbar^2 r_{\mathrm{dd}}}/{m}$ and $-r_{\mathrm{dd}}$ appear in Eqs.~(\ref{aV0Vdd}) and (\ref{adBorn}), respectively.

When the both short-range and dipole-dipole interactions are proportional to a small parameter $\lambda$ \cite{integrablility}, the scattering length can be calculated as the Born series with respect to this parameter
\begin{align}
a_{\mathrm{reg}}=&\, a_{0} - r_{\mathrm{dd}} + a_{1} -\frac{24}{25} \frac{r^2_{\mathrm{dd}}}{r_0}+\cdots, \label{adBornall} \\
a_{0}=& \frac{m}{4\pi\hbar^2}V_{0}(q=0), \label{a0}\\
a_{1}=& -\frac{m}{4\pi\hbar^2}\int\frac{\mathrm{d}^3q}{(2\pi)^3}\frac{V_{0}^2(q)}{2T_{q}}, \label{a1}
\end{align}
where $V_{0}(q)$ is the Fourier transform of the short-range potential. The first two terms in the r.h.s. of Eq.~(\ref{adBornall}) are proportional to $\lambda$ and thus relate the first Born approximation, while the second two terms correspond to the next order of the Born series.

Equation (\ref{adBornall}) can be easily derived from the general relation for the first two terms of the Born series for the scattering amplitude corresponding to the total potential (\ref{fullint}) with the regularized dipole-dipole interaction: ${4\pi\hbar^2a_{\mathrm{reg}}}/{m}=V(0) - \int\mathrm{d}^3q{V^2(\bm{q})}/{[(2\pi)^3 2T_{q}]}$. We find that $V(0)=\frac{4\pi\hbar^2}{m}(a_{0} - r_{\mathrm{dd}})$ and $V(\bm{q})=V_{0}(q)+V_{\mathrm{dd}}(\bm{q},r_0)$ for $q\not=0$, where $V_{\mathrm{dd}}(\bm{q},r_0)$ is given by Eq.~(\ref{vddqr0}). We emphasize that $V(q=0)\not=V(\bm{q}\to 0)$ for the regularized potential.

\section{The Bogoliubov model and the universality of low-density expansions for the dipolar Bose gas}
\label{sec:universality}

There are two kinds of expansions for the Bose gas at zero temperature. The first approach, developed by Bogoliubov \cite{Bogoliubov47}, assumes that the interactions between bosons are weak, that is, the interaction potential is integrable and proportional to a small parameter, coupling constant.  Then it is possible to expand the ground-state energy in this parameter. The second approach \cite{Lee57} is applicable when the potential is strong (say, like the hard-sphere) but the density is small. In this case, the expansion parameter is the density of particles.

The two approaches are closely related. In three dimensions, the density expansions for the strong potentials can be obtained from the expansions with respect to the coupling constant \cite{cherny00}. It is possible if we adopt their \emph{universality}: the system is so dilute that the probability of three-particle collisions is small, and only two-body scattering plays a role. This implies that a few first terms in the low-density expansions can depend only on the \emph{low-momentum scattering amplitude}, which contains two parameters: the scattering length $a$ and the effective range of the dipolar potential $r_{\mathrm{dd}}$ \cite{scat_length}.

Apparently, the idea of universality was first expressed by Lev Landau back in 1947 [see the footnote after equation (30) in the original paper by Bogoliubov \cite{Bogoliubov47}]. A discussion of universal and non-universal effects for Bose gases with short-range potentials can be found in, e.g., Refs.~\cite{Giorgini99,Braaten01}.

The divergence arises, since we ignore the short-range two-particle correlations in real space, and thus the long-range behaviour of the scattering amplitude in momentum space. Therefore, in order to get rid of the divergence, one should develop a consistent scheme taking into consideration the short-range correlations \cite{cherny00}.

However, there is a much simpler way of solving this problem: it is possible to use the universality of the expansion of the energy in the dipolar Bose gas. This can be done by analogy with Sec.~V of the paper \cite{cherny00}. We substitute the Fourier transform of potential (\ref{fullint}) into the Bogoliubov expression for the mean energy (\ref{Boggs}), where $V_{0}$ is supposed to be weak and integrable, and the dipolar interaction is regularized as discussed in Sec.~\ref{sec:Born_reg}. Then in the limit of small coupling constant $\lambda$, we find
\begin{align}
\varepsilon =& \frac{2\pi\hbar^2}{m}\left(a_{0} - r_{\mathrm{dd}} + a_{1} -\frac{24}{25} \frac{r^2_{\mathrm{dd}}}{r_0}\right)n \nonumber\\ &+\frac{256\sqrt{\pi}\hbar^2}{15m}a_{0}^{5/2}{\cal Q}_{5}\left(\frac{r_{\mathrm{dd}}}{a_{0}}\right)n^{3/2}. \label{boggrten}
\end{align}

We are interested  in the first two terms of the expansion, which is assumed to be universal
\begin{align}\label{expeuni}
\varepsilon=c_{1}(a,r_{\mathrm{dd}}) n^{\alpha_{1}} + c_{2}(a,r_{\mathrm{dd}}) n^{\alpha_{2}} +\cdots
\end{align}
with unknown exponents $\alpha_{1}$ and $\alpha_{2}$. The coefficients $c_{1}$ and $c_{2}$ are also unknown functions of $a$ and $r_{\mathrm{dd}}$.

It is possible to find these functions by using the Bogoliubov expansion (\ref{boggrten}). Indeed, the above expansion is valid for arbitrary potential $V_0(r)$. This includes weak potentials, for which the Born series for the scattering length is applicable (see the discussions above in Secs.~\ref{sec:Born} and \ref{sec:Born_reg}): $a=a_{0} + a_{1} -\frac{24}{25} \frac{r^2_{\mathrm{dd}}}{r_0}+\cdots$, where $a_{0}$  and $r_{\mathrm{dd}}$ are proportional to a small parameter $\lambda$, and $a_{1} \sim\lambda^2$. Then we can substitute the Born series for $a$ into the coefficients $c_1$ and $c_2$ and consider the term $a_{1} -\frac{24}{25} \frac{r^2_{\mathrm{dd}}}{r_0}$ as a correction. Expanding the coefficients in the vicinity of the terms proportional to $\lambda$ yields
\begin{align}
\varepsilon= &\left[c_{1}(a_{0},r_{\mathrm{dd}})+ O(\lambda^2)\right] n^{\alpha_{1}} \nonumber\\
&+ \left[c_{2}(a_{0},r_{\mathrm{dd}})+ O(\lambda^2)\right] n^{\alpha_{2}} +\cdots \label{expeuni1}
\end{align}
where $O(\lambda^2)$ denotes the terms of order $\lambda^2$ or higher. By comparing this equation with the Bogoliubov expansion (\ref{boggrten}), we can easily understand how the correct expansion (\ref{expeuni}) looks: we should just replace $a_{0}$ by $a$, completely neglect the term $a_1-\frac{24}{25} \frac{r^2_{\mathrm{dd}}}{r_0}$, and put $\alpha_{1}=1$ and $\alpha_{2}=3/2$. We finally arrive at the correct formula (\ref{expen}).

This reveals the nature of the divergence that we neglect while deriving this formula in Sec.~\ref{sec:bog}: it arises from the second-order terms in the Born series with the pseudopotential (\ref{Vpseudo}). As discussed above, one can completely neglect these terms, even if they are divergent. It is the universality that ensures the correctness of this trick.

\section{Conclusion}

In this work, we studied the expansions of energy (\ref{expen}), chemical potential (\ref{mudipolar}), and condensate depletion (\ref{expdepl}) at small densities. The long-range dipole-dipole potential is regularized with the appropriate regularization, which leads to the jump of the effective scattering amplitude at zero momentum. This procedure allows us to obtain the leading term $-4\pi/3 d^2 n$ in the expansion of the chemical potential.
This term has a simple physical interpretation: it is the classical energy of a dipole in the local electric or magnetic field created by distant dipoles encircling the dipole. The term gives the thermodynamic instability at $\epsilon_{\mathrm{dd}}=1$, which is exactly the same value of the ratio when the Bogoliubov spectrum becomes complex (the dynamic instability). However, the Lee-Huang-Yang correction can stabilize the system for $\epsilon_{\mathrm{dd}}>1$ at the price of breaking the translation invariance and forming droplets.

The asymptotics of the normal (\ref{normas}) and anomalous (\ref{anomas}) correlators and the pair distribution function (\ref{gras}) are calculated analytically without using the regularization of the dipole-dipole potential. As the dipolar forces get larger, the decaying parts of the correlators strongly depend on the angle between the dipoles and the relative distance. When $\epsilon_{\mathrm{dd}}$ exceeds some critical value, the decaying part changes sign with increasing the angle. The pair distribution function, which  gives the probability to find a particle near another particle, is located within the very narrow cones near $\theta=0$ and $\theta=\pi$ in the vicinity of $\epsilon_{\mathrm{dd}}=1$. This means that the dipolar gas should have some tendency to form filaments along the dipole direction when the dipolar ratio is close to one.

We discuss the nature of the divergence arising in the calculation of the energy within the Bogoliubov theory with the effective potential. It is shown how to use the universality of the expansions to avoid the divergence. As a by-product, we consider the two-body scattering problem with zero relative momentum for the dipole-dipole interaction and derive the low-momentum scattering amplitude (\ref{uqeffps}), the asymptotics of the wave function (\ref{phiasym}), and the correction to the scattering length (\ref{adBorn}) for small values of the dipolar range.

The suggested regularization of the dipolar interactions gives quite a consistent picture. Nevertheless, it is worth comparing the results of this paper with Monte Carlo simulations, which have not been done yet for a homogeneous dipolar gas to the best of the author's knowledge. The obtained density expansion for the ground-state energy can be useful for constructing the local density approximation.

\section{Acknowledgement}

The author acknowledges support from the JINR--IFIN-HH projects and thanks the hospitality of the IBS Center for Theoretical Physics of Complex Systems. The author is grateful to Mikhail Fistul and Alexander Chudnovskiy for useful comments on the manuscript.

\bibliography{dipol3D}

\begin{thebibliography}{38}
\bibitem [{div()}]{divergence}%
  \BibitemOpen
  \href@noop {} {}\bibinfo {note} {Note that the integral diverges at large
  distances. It should be multiplied by $\exp(-\lambda q)$, and finally, the
  limit $\lambda\to 0$ is taken.}\BibitemShut {Stop}%
\bibitem [{int()}]{integrablility}%
  \BibitemOpen
  \href@noop {} {}\bibinfo {note} {The potential $V_{0}(r)$ must be integrable
  for the existense of the Born series.}\BibitemShut {Stop}%
\end{thebibliography}

\end{document}